\documentclass[11pt]{article}
\usepackage[margin=1in]{geometry}
\usepackage[T1]{fontenc}
\usepackage{graphicx}
\usepackage{booktabs}
\usepackage{array}
\usepackage{amsmath}
\usepackage{amssymb}
\usepackage{newpxtext,newpxmath}
\usepackage{microtype}
\usepackage[numbers,sort&compress]{natbib}
\usepackage{caption}
\usepackage{url}

\title{AI with Authority, from Application to Silicon}
\author{Jason Hickey (corresponding author)\\[2pt]\texttt{jason@karyk.com}}
\date{}

\begin{document}

\maketitle

\begin{abstract}
For sixty years, machine verification has been a major cost overhead, affordable only
for exceptional artifacts. Here we report that generative AI inverts this relationship:
at AI speed, machine verification is not only economical but essential to productivity ---
it is the incorruptible referee that lets one person safely direct autonomous machine
work at scale. In five weeks, one researcher on consumer AI subscriptions directed a small fleet
of AI agents from application code, through a verified compiler and executive, to a
RISC-V processor taped out on a community silicon shuttle; no proof passed
through human review, and no RTL was written by a human. The working discipline --- the
Salt method --- rests on a proof kernel no hallucinated proof can pass: mathematical claims
travel between agents as kernel-checked artifacts, and human attention is reserved for
statements, designs, and rulings. Verification is stated link by link, from the Lean 4
kernel to SAT-checked equivalence at the silicon boundary. We publish the complete
accounting: theorem provenance, a pre-registered token meter, floor-bounded human time,
and an error ledger whose catch numbering runs to \#256 --- a monotone counter over the
mathematics campaign's append-only flags ledger, maintained 2026-07-07 to 2026-07-20
(one number, \#79, was never assigned; later catches are recorded un-numbered) ---
against zero incorrect proofs reaching the record.
\end{abstract}

\noindent\emph{Author's note: corpus and size counts were extracted at a single commit on
2026-08-14 (the extraction record is published in the mathematics repository); the
priority survey ran 2026-08-11. All kernel-checked claims are current as of this
version.}

\section*{1. Introduction --- a personal journey}

Machine verification is as old as programming, and for most of its sixty years it has
been an exceptional undertaking. When verification is performed by humans, with methods
from Floyd--Hoare logic to modern proof kernels \citep{floyd1967meanings,hoare1969axiomatic}, formal assurance is costly and
impractical. It is affordable for landmark artifacts and little else. Furthermore,
when requirements change, much of the verification must be redone. Generative AI has changed the cost of producing
candidate proofs and designs; it has not, by itself, changed the cost of trusting them.
This paper is a case study in closing that gap --- and, for its author, it is the closing
of a forty-year loop.

The author began this work in 1985, designing switch fabrics for ATM networks at Bell
Communications Research; in 1990 he published a theorem about self-routing switching
networks (ISS '90) \citep{hickey1990iss,marcus1990isscc,marcus1990jssc}. By 1992 he had concluded that software development --- even for hardware ---
was the central bottleneck, and went to study at Cornell University with a mission: to build AI that
develops software. The AI of that era was not equal to the task, and the mission was
deferred --- not abandoned --- while the author spent the next fifteen years, in graduate school
and then as an Assistant Professor at Caltech, building the other half: formal reasoning
systems, including the MetaPRL proof system \citep{hickey2003metaprl}. The mission then waited another
fifteen years for the AI to arrive. Now, in 2026 --- forty years after the work began --- the
two halves have met: the 1990 theorem is proved in a proof kernel and the network taped out
on a community silicon shuttle, designed and submitted in weeks, in evenings and
weekends, by one person directing a fleet of AI agents. The instrument that closed this
loop --- an AI fleet grounded in formal methods --- is the subject of this paper. It is the
machine the 1992 mission described.

This is a case study of that method, which we call the Salt method. Verified stacks are not new: the
field's landmarks --- the CLI verified stack, CompCert, seL4, CakeML
\citep{bevier1989systems,leroy2009compcert,klein2009sel4,kumar2014cakeml} --- were built by expert
teams over years; that lineage is our baseline. What we demonstrate here is
that the economics is efficient. A single person can develop a system stack, each
layer verified at a grade the paper states exactly, from application to silicon tapeout
using consumer products in five weeks --- starting cold: both repositories begin from empty trees atop the public mathlib
library, first commit 2026-07-06 (\S7). It is, we believe, the most completely documented
instance of AI-assisted research and engineering conducted under machine verification ---
documented in the sense that every mathematical claim is kernel-checked and replayable,
every registered theorem carries a commit-verified provenance trail, the economics were
metered, and the errors are recorded in append-only logs. We make no claim that this configuration is optimal, typical,
or generalizable to other researchers; it is a single person, and the author is a
formal-methods specialist. What the case study establishes is an existence proof with measurements
attached.

The context makes the measurements timely. Industrial AI systems now ship complete formal
artifacts: a 56-author Nature paper on olympiad-grade formal reasoning \citep{alphaproof2025}; a frontier
autoformalization agent producing a Riemann-hypothesis-for-curves development \citep{mathinc2026rhcurves}; a new bound on a
longstanding zeta-function problem --- obtained autonomously by an AI system, formalized in
Lean, externally reviewed, and published by its lab the day before this manuscript was first drafted \citep{anthropic2026zeta,claude2026zeta23};
a 91,000-line verified prime-gaps library \citep{axiommath2026primegaps}; a 986,000-line LLM-generated economics corpus \citep{garg2026econcslib}. The field's
question has shifted from whether AI can produce verified mathematics to what it costs, who
can afford it, and how its reliability is governed.

This paper contributes: (1) \textbf{the Salt method}, a framework for highly autonomous AI
development grounded in formal methods --- its required invariants and reference
configuration are stated in full; (2) \textbf{its demonstration at full stack} --- one person, five
weeks, consumer subscriptions --- from application to silicon tapeout, each layer verified
at a grade the paper states exactly; and (3) \textbf{the complete measured accounting} of that
demonstration: the economics under a pre-registered instrument, the error ledger, and the
limits. The central finding is economic: at AI speed, machine verification is not only
economical but essential to productivity --- the sixty-year premise of verification as a
cost overhead inverts, and the kernel becomes what makes AI-scale development usable at
all.

\section*{Results}

\subsection*{2. Salt --- the methodology}

Trust in a verified artifact has two hard halves. The familiar half is the
implementation: code is hard to read, and machine-checked proof addresses exactly that.
The less familiar half is the specification: a formal statement is also hard to read,
and a proof is only as good as the statement it proves. The Salt method is built around
both halves.

The workflow has one shape at every scale: the human prompts an agent in English, and the
agent returns five artifacts --- an implementation; a specification; a machine-checked proof
that the implementation meets the specification; tests, including adversarial controls that
check the proof has bite; and formal certificates that restate the specification in
simplified vocabulary, so that a reader can comprehend what was proved without trusting the
full formal development. The last artifact deserves emphasis: a proof is only as good as
the statement it proves, and the certificate layer is what makes statement-level review ---
the one duty that remains human --- tractable. The certificate's contract is implication,
kernel-checked --- the proved statement entails the restatement --- so a reader can only be
reading something weaker than what was proved, never stronger. Its most familiar form
is a test: for the working engineer the experience is routine --- write tests as always,
with one extra operation, promoting a test to a theorem. The spine's per-round kernel
fixtures (\S5) are exactly such promotions.

The certificates are the entry point to comprehension, not its end: review in this
workflow is \textbf{active interrogation}. 
The human questions the system, and the system answers with kernel-checked
evidence --- unfolding a statement's binders, justifying a hypothesis, producing a variant
under a changed assumption. The author works this way daily; the campaign registers are
substantially a transcript of it. A referee can do the same with the public artifact.

The referee, however, has two ends it structurally cannot check, and the method
is explicit about both. At the back sits the certificate layer just described:
whether a proved statement means what its reader thinks it means. At the front
sits the question verification is most often accused of merely relocating:
whether the specification is what the human wants. That correspondence lives in
the human alone, and the method works it as a discipline rather than an
assumption. The process begins by writing down the objective in prose.  That
objective then receives a structured adversarial review: inconsistencies,
completeness, the population covered, the negative space deliberately
excluded, and suggestions ranked with benefits and downsides. A recorded
interview follows, in which the human's answers are the ground truth being
elicited. The revision is a full set of consistent requirements --- still
prose --- that becomes the pre-registered source for specification and implementation,
with acceptance criteria registered before work begins, so success is never
redefined after the fact. A specification written after the code is a
description; written before, it is a requirement. The human's irreducible
authority thus sits at exactly the two human-language ends of the pipeline ---
saying what is wanted, and reading what was proved --- and the method's whole
project is making everything between them run under the referee. The elicitation
practice predates this campaign, in the author's prior agent collaboration.

Stated concisely, the method makes three commitments. First, truth is machine-checked
only: every claim lands in the kernel, and nothing unverified accumulates into the
record. Second, whatever the kernel cannot check is checked by structured opposition:
designs receive adversarial refuter passes before execution, landings are witnessed
independently, measurements travel with the commands that produced them, and every
control must be able to fail. Third, human attention is treated as the scarcest
resource in the system: it is spent exclusively on statements, designs, and rulings;
tasks are classified by difficulty and priced before they are attempted; and an agent
that exhausts its budget stops and announces its failure rather than grinding on.

In full, the method has six required invariants and six advisory articles; the required
tier is tool-agnostic, and the advisory tier is the reference configuration this case
study ran and measured. The six required invariants are: (R1) the five artifacts above, at every level from
project design to component design; (R2) no claim is admitted without its checker --- the
kernel for mathematics, the named instrument for measurements, structured opposition
for designs; (R3) all design decisions, including human choices, are recorded in an
append-only ledger whose distinctive content is the errors and retractions, amended at
their source and recorded as first-class results; (R4) no statement is ever weakened to
admit a proof --- statement changes are design-tier acts, never taken by an executor;
(R5) a small class of irreversible, outward-facing acts is reserved to human hands, and
the system's job is to reduce each to a prepared click and stop; (R6) conditional
objectives are allowed --- a statement may name hypotheses it does not discharge,
provided each is named in the statement itself and carries a declared disposition,
either \emph{to be discharged} (ledger-owed, with the expectation of a future kernel proof)
or \emph{out of domain} (a stated trust boundary with another discipline, such as
semiconductor physics); a program's final deliverable carries no undischarged in-domain
hypotheses. The six advisory articles, as run here, are: (A1) a single master orchestrator on the top model
class performs the most complex design work, passes routine work to executors, and owns
the referee's own infrastructure --- audit tooling is never owned by a seat it audits;
(A2) executors are the workhorses --- building, verifying, proving, refuting --- and every
task is classified by difficulty and priced before it is attempted;
(A3) attempts are budgeted small, and an agent that exhausts its budget stops and announces its failure
rather than grinding on;
(A4) every major design phase has an exploration part and an adversarial refutation
part, iterated until dry, with acceptance criteria pre-registered before the artifact
exists; (A5) human interaction is periodic and scheduled --- this program held a daily
council with recorded rulings; (A6) every landing is verified by a second agent that
did not produce it. Figure~1 states the method in brief.

\begin{figure}[htbp]
\centering
\includegraphics[width=\textwidth]{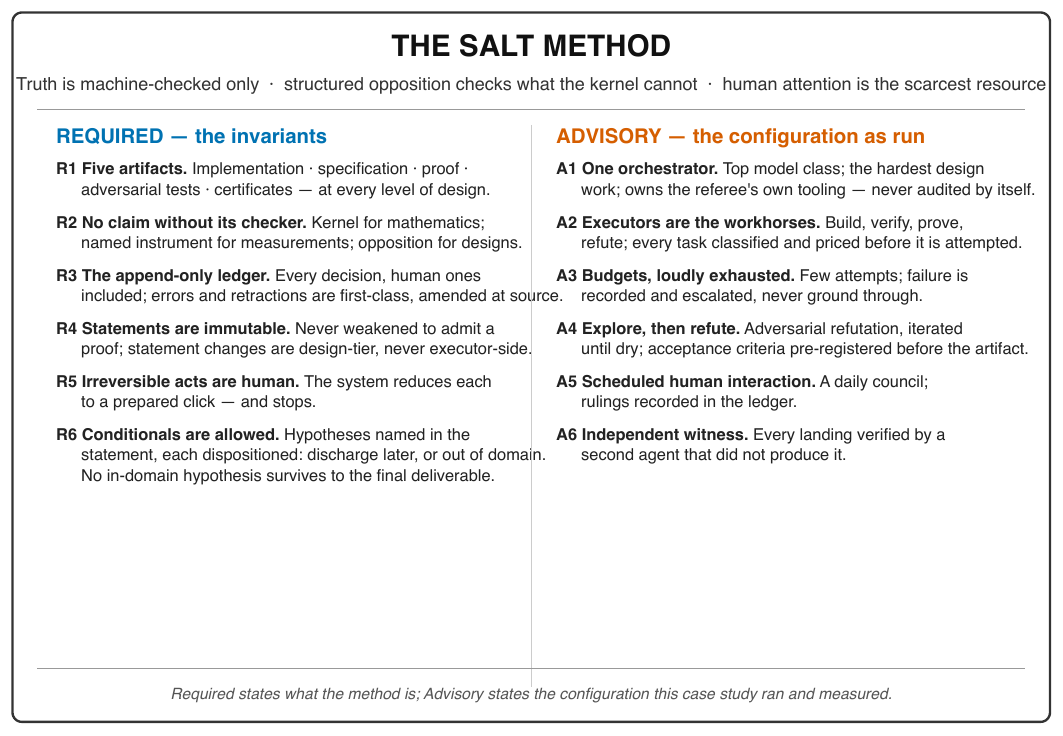}
\caption{\textbf{Figure 1 | The Salt method, in brief.} The creed and the twelve articles --- six required invariants and six advisory articles of the reference configuration. Required states what the method is; Advisory states what this case study ran and measured. The figure is designed to stand alone.}
\end{figure}

\subsection*{3. The referee --- the ground it stands on}

Every mathematical claim in this work is checked by the Lean 4 kernel against mathlib
\citep{demoura2021lean4,mathlib2020}
(version pins in Methods), with per-theorem axiom audits (the three standard axioms only;
no \texttt{native\_decide}; no custom axioms). The kernel is the sole arbiter of mathematical truth
in the workflow: no proof is reviewed by the human, ever. The trust base is
deliberately small: every proof is re-checked by a compact, independent kernel in the
LCF tradition (the de Bruijn criterion), so trust rests on that kernel and on the
statement, never on the model that produced the proof or the fluency of its
explanation. Review collapses to two
questions --- did it check, and is the \emph{statement} the intended one.

The hardware side is checked by a chain of three independent checkers, and we disclose its
structure exactly because it is not uniform. Link 1 --- from specification to emitted design
artifacts --- is kernel-checked in Lean. Links 2 and 4 --- that the Verilog the toolchain
consumes corresponds to the emitted artifacts, and that the synthesized netlist corresponds
to that Verilog --- are SAT-based equivalence checks (Yosys), and can only be that: as of 2026-08-11, no
general Verilog-to-Lean importer exists in any public artifact (the campaign's own
importer, \S5, is scoped to this flow's netlist-level Verilog, not general Verilog), so a
kernel cannot referee those links today. Link 3 is the synthesis miter. The chain's three checkers were built by three different
agents, disagreed twice during the campaign, and were reconciled at the byte level; the
disagreements are in the ledger. We regard the non-uniformity of this chain as a finding,
not a flaw: it maps exactly where today's verified-hardware boundary sits for a small team.

\subsection*{4. The fleet --- the methodology in practice}

The configuration for this case study consists of five long-running AI agent seats --- a
coordinator, mathematics, compiler, silicon, and evidence --- sharing a repository and an
append-only message bus, all directed by one human (Fig.~2). The seats operate under written laws
that exist because the kernel's ground truth makes them enforceable. Every landing is
independently witnessed by a second seat. Designs receive adversarial refuter passes before
execution consumes them. Measurement precedes assertion, and the extractor command that
produced a number travels with the number. Errors are amended at their source rather than
corrected downstream. Every attempt carries a budget after which the executor must stop and
announce its failure, and failures are recorded in a ledger alongside the results.

\begin{figure}[htbp]
\centering
\includegraphics[width=\textwidth]{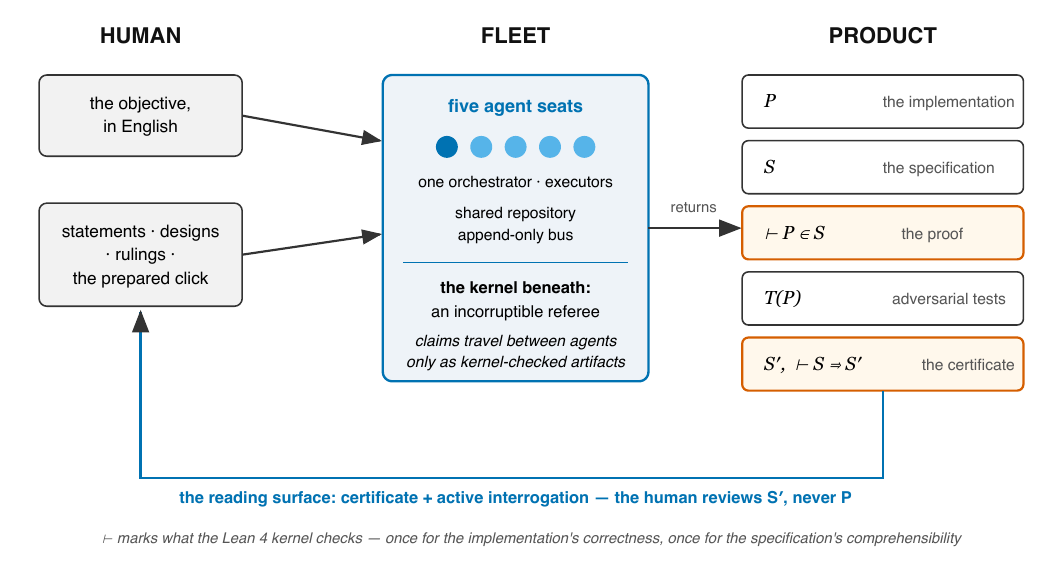}
\caption{\textbf{Figure 2 | The configuration and its product.} One human directs a five-seat agent fleet sharing a repository and an append-only message bus, with the kernel beneath as the incorruptible referee; every objective returns five artifacts --- the implementation P, the specification S, the kernel-checked proof $\vdash P \in S$, adversarial tests T(P), and the certificate $S'$ with $\vdash S \Rightarrow S'$ kernel-checked. The return arrow is the method's reading surface: the human reviews the certificate and interrogates; no proof passes through human review. $\vdash$ marks what the Lean 4 kernel checks --- once for the implementation's correctness, once for the specification's comprehensibility.}
\end{figure}

The observable consequence is an error record; the campaign-wide single ledger is
constituted at the pre-publication freeze
(the extractor is designed; its unit is the incident, never the mention). Over the mathematics campaign
the system's adversarial layers --- the kernel first, then the seats' cross-checks --- caught
design errors on a catch ledger whose numbering runs to \#256 --- a monotone counter over
the append-only flags ledger, maintained 2026-07-07 to 2026-07-20; \#79 was never
assigned, and later catches are recorded un-numbered and excluded ---
among them wrong scope on a measured claim, stale citations,
misattributed mechanisms, and statement-level
type traps. In the same period, zero incorrect proofs reached the record (the kernel makes
this class structurally impossible to record), and the register's integrity is itself
machine-checkable: all 73 registered headline theorems carry stated landing dates matching
their landing commits, and all 59 landing commits are ancestors of the main branch. The
ledger's composition is data about AI-assisted research: a hand-classified breakdown
exists for the first 78 numbered catches (classified 2026-07-15); the later catches are
unclassified, and no class-dominance claim is made here. The
workflow's culture of same-day retraction at the source is, we will argue in \S8, the
referee's most important export beyond the proofs themselves.

\subsection*{5. The spine --- the demonstration, from application to silicon tapeout}

The paper's central artifact is a systems stack with verification stated link by link
(\S3), built in seven days of elapsed
repository history, inside the program's five weeks: a compiler from a structured language
to a small instruction set, with simulation proofs for its control constructs; a
multitasking executive; and the silicon design. The design was first submitted 2026-08-10 to
Tiny Tapeout's September 7, 2026 community shuttle and revised before shuttle close; the
shipped design of record is the revised submission (shuttle run 32284710003,
shuttle-repository commit \texttt{7d2b275}). The provenance census is stated for the
2026-08-10 submission, the design the structural join has measured: it carried 902
flip-flops of sequential state, of which
288 (31.9\%) were emitted from kernel-checked Lean artifacts and 614 were from agent-written RTL.
A fourth kernel-emitted MAC island (64 flip-flops in RTL) was deliberately instantiated
disabled and correctly removed by synthesis --- measured at the GDS by the structural join,
which reaches all 288 named flip-flops with zero misattributions; at the RTL the emission count
was 352 of 966 (36.4\%). Both scopes are stated because they answer different questions;
the scope sentence travels with every telling of this number. In the shipped revision the
kernel-emitted RTL re-derives exactly --- four MAC islands and three serializers, 352
kernel-emitted flip-flops instantiated, the fourth island again tied disabled --- and the
committed synthesis stat records 1,468 sequential cells in total; the structural join has
not yet been re-run on the shipped run's GDS, so no die-level provenance ratio is stated
for the shipped design. The 1990 switching-network
theorem rides on the design: the routing schedule it certifies is proved in the kernel (the
full rotation-closure result) and drives the submitted switch (Fig.~3).

\begin{figure}[htbp]
\centering
\includegraphics[width=0.55\textwidth]{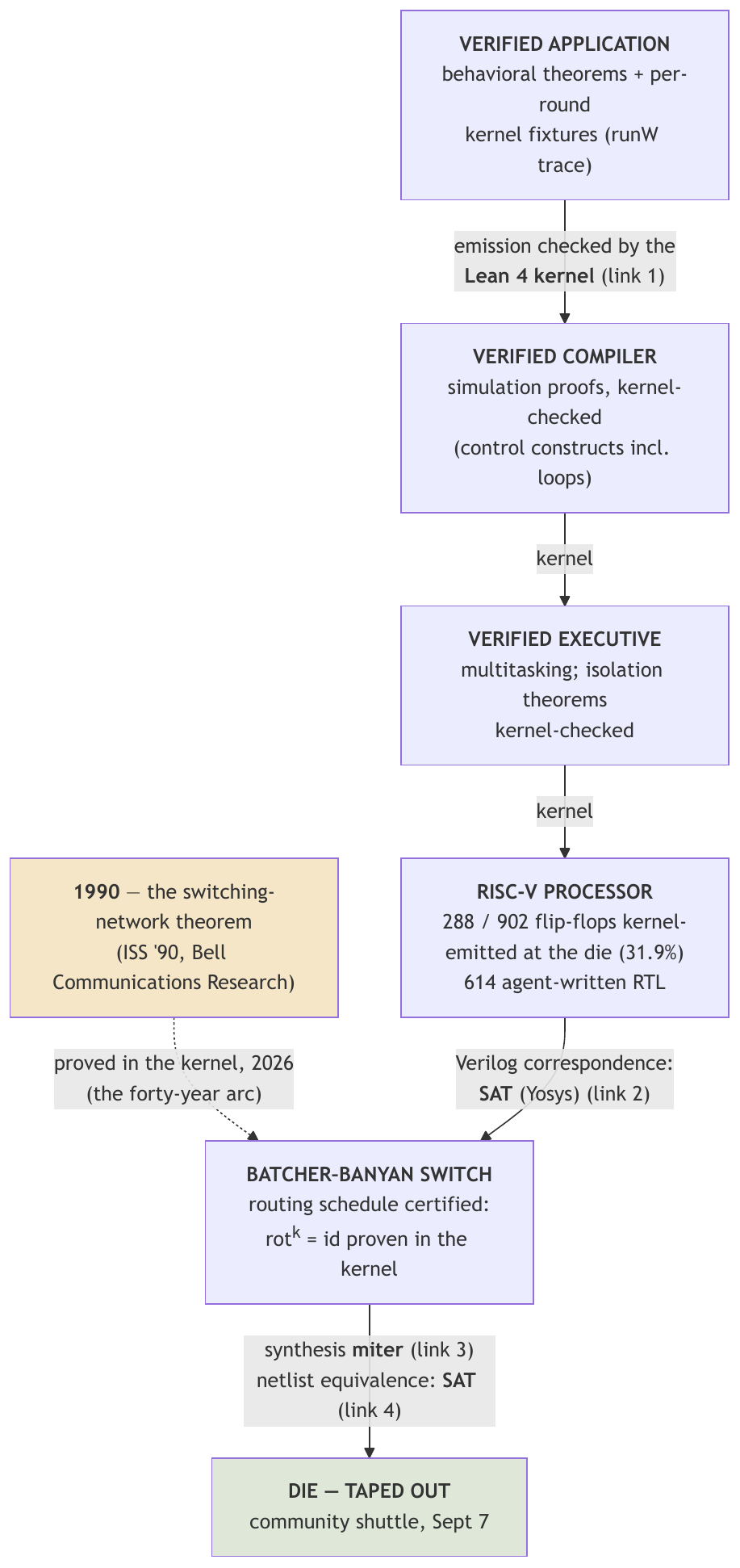}
\caption{\textbf{Figure 3 | The spine as a chain of custody.} The four parts and the links between them, each link labeled by its checker (Lean kernel; SAT equivalence; synthesis miter). The processor's sequential state is shown at its measured provenance in the 2026-08-10 submission (288 of 902 flip-flops kernel-emitted, 31.9\%; RTL-side 352 of 966 --- the disabled fourth MAC island was correctly removed by synthesis); the shipped revision's GDS has not yet received its structural join, and no die-level ratio is stated for it. The accompanying table gives each component's measured size --- sizes to be drawn beside the components in the final figure. The arc above traces the 1990 switching-network theorem from its publication to its kernel proof and its place on the submitted design.}
\end{figure}

\begin{table}[htbp]
\centering
\begin{tabular}{@{}>{\raggedright\arraybackslash}p{0.30\textwidth}>{\raggedright\arraybackslash}p{0.60\textwidth}@{}}
\toprule
Component & Size (measured; extractors in Methods) \\
\midrule
Verified compiler (DSL $\to$ ISA) & 5,067 Lean lines / 13 files (figure retired by \texttt{docs/methods-size-manifest.md} in the systems repo --- the manifest's file list is normative; the row re-derives from it at one sha) \\
Verified executive + application stack & 11,001 Lean lines \\
Silicon flow: importer, equivalence, cell models & 4,251 Lean lines \\
Certificates (systems side) & 1,884 Lean lines \\
Agent-written RTL & 22,679 Verilog lines across 71 files (silicon's measured split, saltworks 456f508 --- denominator reproduced first at 316,911/119; flow-generated netlists 294,232 lines/48 files excluded, 92.8\%) \\
\bottomrule
\end{tabular}
\end{table}


We report the spine not as a hardware contribution --- the design is modest --- but as the case
study's demonstration that one configuration can carry a \emph{single chain of custody} from a
theorem statement, through a verified compiler, to a taped-out physical design (Fig.~4), with the
trust boundaries of \S3 named at each link.

\begin{figure}[htbp]
\centering
\includegraphics[width=0.8\textwidth]{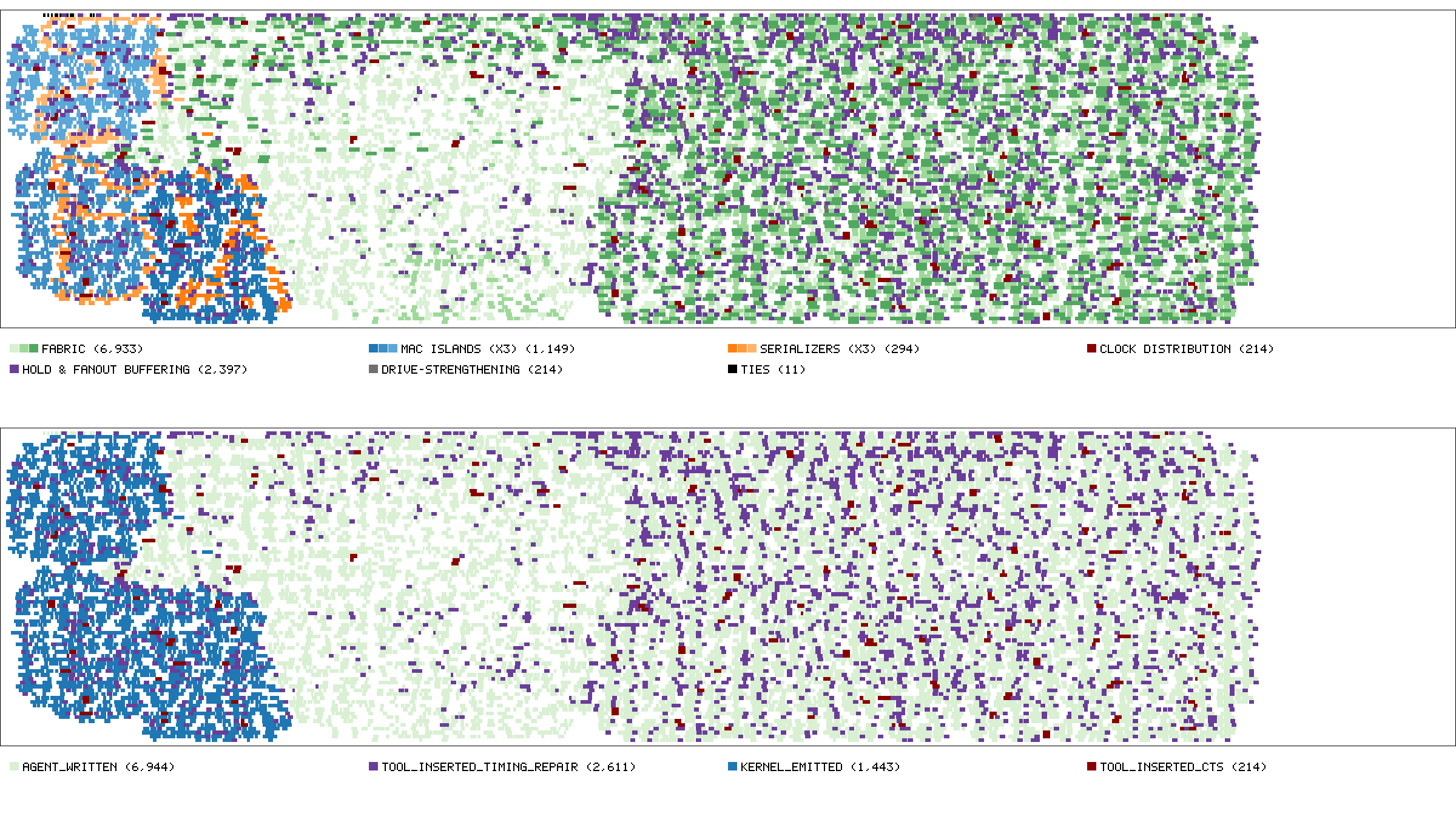}
\caption{\textbf{Figure 4 | The shipped die, logic visible --- two colorings.} The submitted design (Tiny Tapeout shuttle run 32284710003, shuttle-repository commit \texttt{7d2b275}), drawn from the shuttle's own final placement. 43,884 placed instances occupy a $1030.40 \times 225.76\,\mu$m $6\times2$ tile at 56.27\% design-instance utilization (LibreLane). 11,212 are logic and are drawn; the remaining 32,672 are fill, decap, tap and antenna cells and are not drawn, which is what the white space is; the $6\times2$ die area is outlined, so the placed logic is seen against the true die edge. \textbf{Top, function.} Color follows function, and shade follows the member within a family: the fabric (including the RISC-V core and control) --- 6,933: 3,739 combinational, 2,014 multiplexer, 1,180 sequential; the three MAC islands (1,149, kernel-emitted by the Lean-verified emitter, not ``kernel-verified''), each physically interleaved with its serializer (294), which is why per-function boxes would overlap and per-cell coloring is the faithful form; clock distribution (214); hold \& fanout buffering (2,397 --- 1,267 hold-fix delay cells and 874 max-fanout buffers on data paths, plus 256 slew, capacitance and wire-length repairs); drive-strengthening (214); and ties (11). Of the die's 1,468 flip-flops, 288 lie inside the named MAC and serializer groups and 1,180 in the fabric. \textbf{Bottom, provenance} --- who authored each cell: agent-written RTL 6,944 (61.9\%); tool-inserted timing repair 2,611 (23.3\%); kernel-emitted 1,443 (12.9\%); and clock distribution inserted by clock-tree synthesis 214 (1.9\%), which is authored by nobody --- a category as distinct from agent-written as from kernel-emitted. 1,443 cells (12.9\% of logic) carry a surviving hierarchical name; the rest are anonymous after flattening.}
\end{figure}

\subsection*{6. The forge --- the mathematical foundations}

The mathematics was also the forge of the method itself. Each of the method's rules
(\S2) was minted from a practical failure --- hallucinated results, plausible-but-wrong
designs, measurements quoted beyond their scope --- and the ledger records the incident
behind every law: the method was not designed in advance and then applied, but
accumulated as case law under the referee, which is why its articles read like a record
of things that actually went wrong.

The choice of mathematical foundation was, in some sense, accidental. The author set
out to study the twin prime conjecture, and the method condensed out of that campaign
because hard mathematics under a kernel is an unforgiving proving ground. Nothing in the method requires it: we are not suggesting that one must work on
twin primes before designing a chip. Any domain that pairs fast generation with an
incorruptible checker could have forged the same laws; this one happened to be ours.

The corpus this forge produced is the largest measured dataset in the study: over 320,000
lines of Lean 4 under the paper's strict extractor (registered in the repository
record; 658,103 lines by raw count, and both counting methods publish), produced in 37 days --- for calibration, 29.3\% of the size of mathlib itself,
measured with the identical extractor at the pinned revision (Fig.~5).

\begin{figure}[htbp]
\centering
\includegraphics[width=\textwidth]{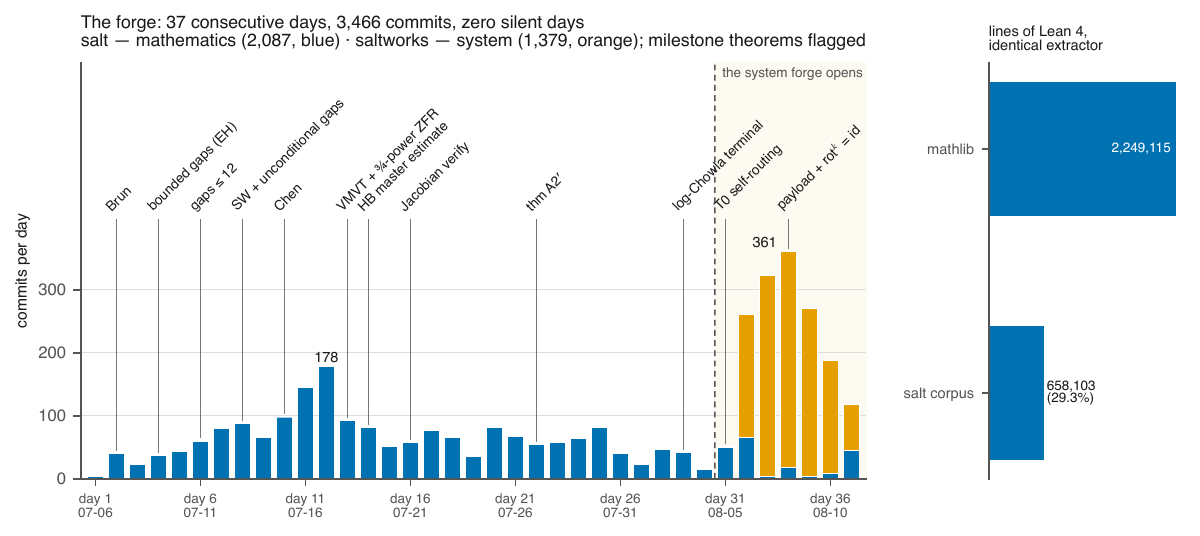}
\caption{\textbf{Figure 5 | The forge.} Commits per day over the campaign's 37 consecutive days --- zero silent days --- stacked by repository: salt (mathematics, blue) and saltworks (system, orange), the dividing line at the system forge's opening (Aug 5); milestone theorems flagged at their landing days. Inset: the corpus against mathlib, measured with the identical extractor at the pinned revision (29.3\%).}
\end{figure}

We state its contents at surveyed strength \citep{saltsurvey2026}: as of 2026-08-11, no public artifact in any
proof assistant proves the Siegel--Walfisz theorem, the large sieve inequality
\citep{montgomery2007multiplicative},
Bombieri--Vinogradov, a lower-bound (Rosser--Iwaniec) sieve, Chen's theorem
\citep{chen1973representation}, the Vinogradov
mean value theorem, the Weil bound for Kloosterman sums, a zero-free region beyond de la
Vall{\'e}e Poussin strength, or Matom{\"a}ki--Radziwi{\l}{\l}/Hal{\'a}sz-type machinery
\citep{matomaki2016multiplicative} --- indeed a live
external Bombieri--Vinogradov formalization project \citep{mellendijk2026bv} takes
Siegel--Walfisz and the large sieve as named axioms in its own source --- and this corpus
carries machine-checked proofs of all
of them, dated 2026-07 on a repository that was private until 2026-08-16, when it was
made public (\texttt{github.com/jyh/salt}). Several
other results were formalized independently of near-simultaneous public artifacts
(Vaughan's identity \citep{mellendijk2026bv}; the Maynard--Tao sieve
\citep{axiommath2026primegaps,maynard2015small}; the Montgomery--Vaughan Hilbert inequality
\citep{claude2026zeta23,montgomery1974hilbert} ---
external artifacts cited), which we report as independent formalizations, not firsts; the
survey method and per-claim evidence are published with this paper.

The program's ambition was the twin prime conjecture, and we state its outcome plainly:
the conjecture remains exactly what it was --- in the corpus it is a definition, never a
theorem, and every conditional result names its hypotheses. What the campaign produced
instead is, we believe, more interesting as a case-study artifact: machine-checked theorems
delimiting the method's own reach. The corpus proves, in the kernel, that no weight in the relevant Maynard-class can cross
the twin gate ($M_2 \le 2 \log 2 < 2$) and that the least $k$ with $M_k > 2$ is five
\citep{maynard2015small,polymath2014variants}, and it proves
a formal gap theorem for parity-invariant sieve certificates. The
fleet aimed at the hardest problem, landed the classical pillars on the way, and then
\emph{verified the wall} --- converting a folklore obstruction into kernel objects. A research
program that can machine-check the boundary of its own methods is, to our knowledge,
without precedent, and it is a capability the configuration gets specifically from the
referee: a barrier argument is exactly the kind of subtle claim that benefits from a
kernel.

\subsection*{7. The economics}

We publish the accounting with its instruments, and we state first what cannot be derived,
because the temptation in this genre is to print ratios the records do not support. From
this project's records one cannot derive a dollar cost per theorem (subscription pricing
carries no per-request prices); model-hours; a per-account attribution; or a
generated-versus-authored split of the Lean corpus. The figures below are
what the records do support.

\textbf{Scale and pace.} The campaign ran 37 consecutive days (2026-07-06 to 2026-08-11) and
produced 2,087 commits in the mathematics repository with zero silent days (mean 56.4
commits/day, peak 178); the systems repository received 1,379 commits over 7 days
(peak 343). Headline results arrived continuously: unconditional bounded prime gaps
\citep{zhang2014bounded,maynard2015small,polymath2014variants} on day 8, Chen's theorem
\citep{chen1973representation} on day 10, the
power zero-free region ($\theta = 3/4$) on day 13; the last headline theorem on day 29.

\textbf{Metered window.} Under a token meter pre-registered before its data accumulated
(instrument and pre-registration published), a 4.86-day window at campaign end recorded
28.07M output tokens across 36,844 deduplicated API requests accompanying 1,376 commits and
56,951 inserted Lean lines --- 376 output tokens per inserted line, a figure whose numerator
includes all prose and design work in the repository and which must not be extrapolated to
the full corpus (the meter postdates every headline theorem; we state this as the study's
largest measurement gap, not a footnote).

\textbf{Human time.} A published-rubric extraction bounds the human's engaged time at
37\,h\,21\,m across 45 blocks, out of the metered window's 116\,h\,40\,m of wall clock, nights
included (the same 4.86 days), with a named uncertainty band of 4 minutes whose authorship
the record cannot settle (excluded and reported, never folded in). The figure was
corrected downward from a first extraction of 44\,h\,25\,m after a cross-seat audit: the
transcript channel carries machine-authored keystrokes --- a coordinating agent nudging
fleet seats by terminal injection arrives with human provenance fields, indistinguishable
at the record layer from a hand at the keyboard --- and correlation against the sending
seat's own transcripts (instrument published) proved 11\,h\,50\,m of such traffic inside the
window, including orders the author explicitly disowned on the record. Because a smaller
human number flatters this paper's thesis, exclusion is the self-serving direction and is
held to machine proof; uncertain cases are excluded and stated as the band. Engagement
blocks bridge gaps up to twenty minutes --- an ordinary phone call counts as engaged --- so
the figure is a coarse envelope of presence, not an attention meter, and no finer
category composition is published at this grain.

\textbf{Unattended operation.} The silence-window instrument behind panel (b), run over the
full campaign, bounds autonomy in both directions. A silence window is the
stretch between consecutive human touches to any personal-lane seat (every agent session
of the author's on this program's side of his employment firewall) --- a claim about
direction, not sleep --- and coverage is disclosed with the run: commits predating the
earliest readable transcript are excluded rather than counted as silent. In the
mathematics repository, 43.0\% of commits landed inside silence windows of at least one
hour, 14.5\% at four hours, and 8.8\% at eight; the longest such window, 20\,h\,56\,m, carried
26 commits and 12,310 inserted Lean lines. The systems repository was driven more
interactively: 24.4\% at one hour, 4.6\% at four, a single commit at eight. Unattended
night operation under standing evening orders was part of the configuration throughout ---
stated in silence-window form because the clock-hour version (18.1\% of mathematics
commits landed 21:00--05:00 local) is the thinner claim, reported once so no reader need
compute it. We state what silence does not mean: the designs being executed were frozen and
refuter-attacked before the window opened; the claim is that the execution loop ran
without direction, not that work appeared from nowhere (Fig.~6).

\begin{figure}[htbp]
\centering
\includegraphics[width=\textwidth]{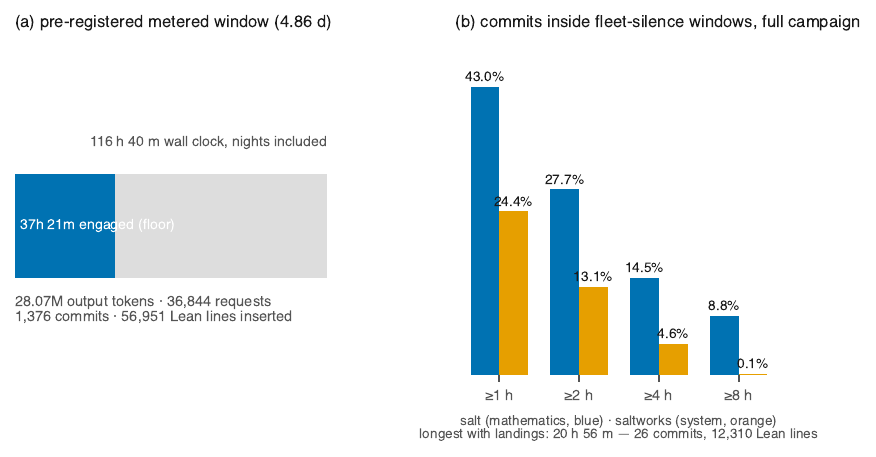}
\caption{\textbf{Figure 6 | The economics, measured.} (\textbf{a}) The pre-registered metered window: 116\,h\,40\,m of wall clock (nights included), the human's engaged floor of 37\,h\,21\,m --- machine-authored keystrokes excluded by the published correlation instrument --- and the window's totals. (\textbf{b}) Share of commits landing inside fleet-silence windows over the full campaign, by repository; the longest silent stretch containing landings ran 20\,h\,56\,m and carried 26 commits.}
\end{figure}

\textbf{The human's role.} The author and the
fleet convened periodically --- typically once a day --- to rule on major design
decisions; between rulings the fleet worked autonomously at the execution layer, no
proof passing through human review. The campaign registers count 20 council sittings with
recorded rulings; 7 irreversible acts (submissions, purchases, sends) taken by the human
against 5 further acts named and deliberately not taken; 1 design veto; 9 source verifications
of the kind only a human with the paper or the vendor portal could perform. The structural
pattern is the finding: authority was \emph{reserved}, not continuously exercised --- the ledger
records agents that reduced a theorem to one click and stopped, by design, because
the click carried the human's word.

{\sloppy 
\textbf{And one retraction, reported as a result.} Mid-campaign, the project measured a
verification-cost ratio, published it internally, and struck it the same day when a second
run --- of three in all --- swung the ratio by a factor of 52 --- one of the three runs falling
inside the very 10--100$\times$ overhead range the claim had denied. The retraction stands in the
ledger (\texttt{5fa8987} $\to$ \texttt{8520580}) with a standing instruction never to quote a ratio of that
class again. We include it
because it is the paper's thesis in miniature: the configuration's value is not that it
produces impressive numbers, but that its numbers are governed.\par}

\textbf{The arc of the role.} In the campaign's first days the author drove everything: each
theorem began as a conversation, model configurations were swapped by hand for every
design run, and he stayed attentive through the nights. Mid-campaign he drove eight hours each way, on a weekend, to visit family ---
laptop tethered to his phone and powered from an oversized battery, pulling off at highway ramps
whenever a theorem finished --- so that no decision would wait on his absence. What changed
over the five weeks was not the amount of his engagement --- the transcripts show it grew ---
but its kind: the machinery he once operated by hand became law-governed and
pre-authorized, decisions moved up the stack from mechanism to statement, and the referee
held the floor in between. By the final week the fleet ran its nights with landings in
his silence, and the author reports the configuration's most personal measurement
himself: he sleeps untroubled. His curiosity has its own category in the pre-registered
rubric --- watching, redirecting nothing --- counted as its own line, proudly.

\subsection*{8. What the referee exports}

The case study's qualitative finding is that the kernel's epistemics leak outward. A fleet
of AI agents whose native failure mode is confident error spent the campaign catching each
other's scope claims, retracting at the source, and converting incidents into written
laws --- because an incorruptible ground truth existed to anchor the culture. The error
ledger shows the classes: measurements published with the scope of laws; registers
asserting world-state instead of measuring it; instruments trusted across configuration
boundaries they were never validated for. Each class was caught, named, and answered with a
mechanism --- by the agents, on the record. On the study's final morning, the fleet's own
priority survey (five adversarial search lanes over the live literature) found that two of the
corpus's believed firsts had public predecessors --- one under a different name that no text
search could see --- and the claims in \S6 are stated at the strength that survey supports. A
workflow that catches its own priority errors before a referee does is the strongest
evidence we can offer for the thesis that verification-grounded process, not model
capability alone, is what makes AI-scale research trustworthy.

\section*{Discussion}

For six decades --- from Floyd and Hoare's program logics forward
\citep{floyd1967meanings,hoare1969axiomatic} --- formal
verification has been priced at a significant multiplier on development cost, with a
significant additional cost whenever requirements change: formal development is not
practical. 
It
is tenable for a compiler, a microkernel, or a landmark theorem
\citep{leroy2009compcert,klein2009sel4,hales2017kepler}, and for little else. The
configuration measured here inverts the sign in a specific regime: when generation is fast,
cheap, and fallible, the kernel is not a tax on production but the precondition for it. One
person can direct work at this scale only because no proof requires their review; their
scarce attention is spent entirely at the statement and design layer, which the campaign's
registers show is exactly where the errors live. We do not claim the inversion holds
outside this regime, and the study's own records show the configuration's edges --- the SAT
links, the autonomy tail (no silence window with landings exceeded 21 hours), the
measurement gaps. What we claim as demonstrated is
narrower and, we believe, of broad interest: machine verification is what turned a fleet of
generative models into a research instrument whose output can be trusted at the
campaign's measured pace --- 2,087 commits in 37 days, zero incorrect proofs reaching the
record (\S\S4, 7) --- and the complete accounting of one such instrument, errors
included, is now public.

\subsection*{Limitations}

This study does not contain new headline mathematics: the twin prime conjecture is
untouched, and the corpus's celebrated theorems are formalizations of known results. The
verified-hardware chain has named SAT-only links (\S3). The subject is one expert
practitioner; nothing here estimates what other researchers, or teams, or other domains
would achieve, and we make no labor-market claims. Priority claims carry as-of dates against
a field moving on a cadence of weeks --- during this paper's own final audit, one competing
library pushed new commits --- and will be re-surveyed at submission. The corpus supports
``present and kernel-checked,'' not ``authored,'' until the generated-versus-authored split is
published. The economics instrument covers the campaign's final window only. Finally, no
physical chip exists yet: the design is a submission to a community shuttle closing
2026-09-07; the vendor's estimated delivery is 2027-05-12 (the shuttle publishes no
fabrication date), and no result in this paper rests on measured silicon.

\section*{Methods}

Fleet architecture (five seats, bus, laws --- full protocol documents published); Lean 4 /
mathlib pins \citep{demoura2021lean4,mathlib2020}; axiom audit protocol; the verification chain per link with tools and
versions; the token meter and its pre-registration; the human-time rubric; the survey
method for \S6's claims (five adversarial search lanes, per-claim evidence files); AI-use
disclosure per journal policy: the agents are Claude-family models (Anthropic) operated
under consumer subscriptions; all agent output was governed as described. This work is
the product of a month-long collaboration between the author and Claude. Text and
figures prepared in this collaboration may carry Anthropic's content-provenance marks
(imperceptible text watermarks; C2PA metadata on image files), consistent with this
disclosure. No AI system is an author, and the author takes full responsibility for the
manuscript. A full model and version enumeration is deferred to the pre-publication
content freeze.

\section*{Data availability}

The mathematics corpus and the systems stack, including the error ledger, have been
public since 2026-08-16 (\texttt{github.com/jyh/salt} and \texttt{github.com/jyh/saltworks});
the remaining campaign registers, including this paper's audit briefs, become public at
publication; every theorem
replays with one command; the submitted design's files
are on the shuttle's public record.

\section*{Author contributions}

J.H. is the sole author: he conceived the work, designed the methodology, directed the
research, and wrote the manuscript, reserving to himself all statements of results,
designs, and rulings, and he takes full responsibility for the originality, accuracy, and
integrity of the work. AI systems are not authors; per the disclosure below and the
journal's AI policy, the Claude-family agents operated as instruments under J.H.'s
direction, and every mathematical claim they produced is machine-checked by the Lean~4
kernel rather than accepted on authority.

\section*{Funding}

This work received no external funding.

\section*{Acknowledgements}

This work was created in collaboration with Claude (Anthropic). The collaboration is
itself the subject of the paper: the agents' contributions --- the proofs, the designs, the
drafts, and the errors caught and corrected --- are documented in the campaign registers
published with this work.


\bibliographystyle{plainnat}
\bibliography{refs}

\end{document}